\documentclass[amsmath,amssymb,11pt]{article}
\usepackage{graphicx}
\usepackage{bm}
\usepackage{amsmath,amssymb}
\usepackage{color}
\usepackage{fullpage}%
\usepackage[affil-it]{authblk}
\usepackage{enumerate}
\usepackage[dvipdfmx,colorlinks,citecolor=red]{hyperref}

\begin{document}
\title{\bf Measurement-free topological protection using dissipative feedback}%

\author[1,2,3]{Keisuke Fujii}
\author[1]{Makoto Negoro}
\author[1]{Nobuyuki Imoto}
\author[1]{Masahiro Kitagawa}

\affil[1]{
Graduate School of Engineering Science, Osaka University,
1-3 Machikaneyama, Toyonaka, Osaka 560-8531, Japan}
\affil[2]{The Hakubi Center for Advanced Research,
Kyoto University, Yoshida-Ushinomiya-cho, Sakyo-ku, Kyoto 606-8302, Japan}
\affil[3]{Graduate School of Informatics,
Kyoto University, Yoshida Honmachi, Sakyo-ku, Kyoto 606-8501, Japan}

        \maketitle  

\begin{abstract}
Protecting quantum information from 
decoherence due to environmental noise
is vital for fault-tolerant quantum computation.
To this end, standard quantum error correction employs
parallel projective measurements of individual particles,
which makes the system extremely complicated.
Here we propose 
measurement-free topological protection in two dimension 
without any selective addressing of individual particles. 
We make use of engineered dissipative
dynamics and feedback operations
to reduce the entropy generated by decoherence
in such a way that quantum information is topologically protected.
We calculate an error threshold,
below which quantum information is protected,
without assuming selective addressing, projective measurements, nor
instantaneous classical processing.
All physical operations are local and translationally invariant,
and no parallel projective measurement is required,
which implies high scalability.
Furthermore, since the engineered dissipative dynamics we utilized
has been well studied in quantum simulation,
the proposed scheme can be a promising route progressing 
from quantum simulation to fault-tolerant quantum information processing.

\end{abstract}

\maketitle
\section*{Introduction}
A standard approach to
protect quantum information from decoherence
is the use of celebrated quantum error correction (QEC)~\cite{Shor95,DiVincenzoShor}.
It conventionally employs projective measurements, 
classical information processing, and feedback operations with selective addressing of individual particles.
Based on the standard paradigm of QEC,
fault-tolerant quantum computer architectures
have been designed with the additional ability to operate universal quantum gates~\cite{Kitaev97,Preskill98,Knill98,AB-O99,Steane,Knill05,Raussendorf06,Raussendorf07b,Raussendorf07a,FY10b,FY10}.

\begin{figure*}
\begin{center}
\includegraphics[width=150mm]{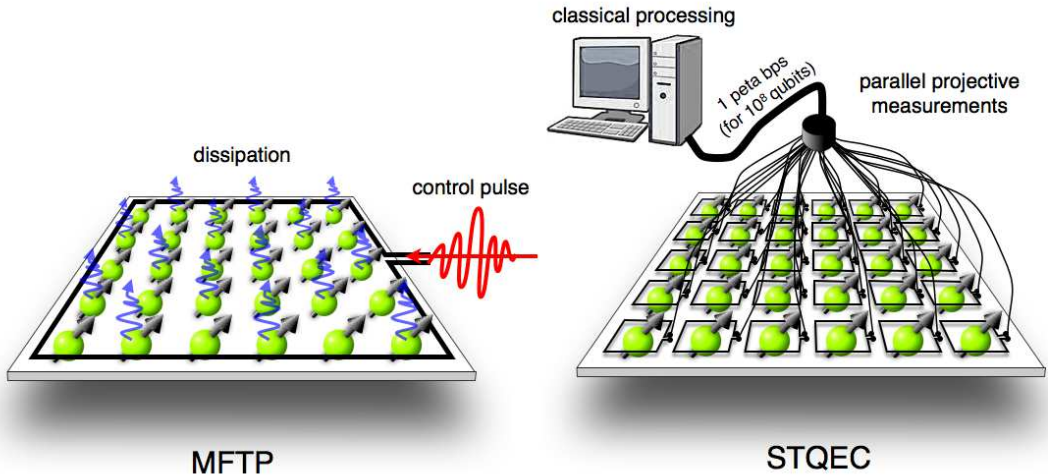}
\caption{MFTP v.s.~standard QEC. MFTP with discrete-dissipative feedback operations,
which are implemented by local and translationally invariant physical operations (left).
A standard QEC with projective measurements and classical processing (right). Since 
MFTP does not require any selective addressing,
its scalability is evident.
}
\label{fig0}
\end{center}
\end{figure*}

However, there exist severe problems that have to be overcome
for the realization of standard QEC.
The projective measurements and classical processing
utilized in standard QEC 
have to be much faster than 
the coherence time of quantum systems,
which is extremely challenging in experiments.
If classical processing depends on the size of the system,
it ultimately limits both the speed and the size of quantum computers.
From a theoretical viewpoint, 
it is highly nontrivial to establish an error threshold theory
including the classical system to control quantum computers.
In practice, fast and reliable parallel projective measurements
of the massive numbers of qubits employed in conventional QEC
are very challenging (see Fig.~\ref{fig0}).
A recent study \cite{Jones} has revealed that
parallel projective measurements of individual $10^8$ qubits 
are required every hundred nsec 
to maintain quantum coherence for factorization of 1024-bit composite numbers.
(The total amount of information measured is 1 peta bits/sec!)
While monolithic architectures,
quantum dots~\cite{Jones} and superconducting qubits~\cite{Fowler,Ghosh} on a chip,
exhibit promising scalability, 
macroscopic measurement devices coupled 
with individual qubits for parallel projective measurements
might introduce other sources of decoherence,
thereby limiting this approach.
On the other hand, distributed architectures, consisting of
modules comprising a small number of qubits, connected with optical channels,
allow both accurate manipulations and measurements inside the local modules~\cite{Distributed,RepeaterQIP,Jiang07,Li10,FT10,FYKI12,Monroe,LiBenjamin}.
However, the entangling operation between separate local modules
using flying photons takes a long time due to photon loss.
Hopefully, these problems will be overcome within the conventional paradigm
by a breakthrough in the development of 
accurate manipulations and parallel projective measurement technology.
In the mean time, we should not stop searching for a novel way toward robust and scalable protection of quantum information.

Here, we propose a new paradigm, measurement-free topological protection (MFTP)
of quantum information using dissipative dynamics, 
paving a novel way toward fault-tolerant quantum computation.
We unify the quantum system that is to be protected
and a controlling classical system in a framework 
without assuming parallel projective measurements, selective addressing, nor instantaneous classical processing.
We restrict our controllability to local and translationally invariant physical operations in a two-dimensional (2D) system.
Since this level of control does not require 
any selective addressing or parallel projective measurements,
MFTP enables us to easily achieve
scalability in various physical systems (see Fig.~\ref{fig0}).
Furthermore, 
engineered dissipative dynamics utilized for topological protection
has been well studied in the context of quantum simulation~\cite{Weimer,Barreiro,Lanyon,Muller}.
MFTP serves as a promising route to progress from
quantum simulation to fault-tolerant quantum information processing.

\section*{Topological protection}
We consider a 2D many-body quantum system 
which encodes quantum information, for simplicity, using the surface code~\cite{Kitaev,Dennis} (the lower layer in Fig.~\ref{fig2s} (a)). 
The proposed scheme can also be applied straightforwardly
to other local stabilizer codes such as topological color codes~\cite{Bombin2}.
The surface code, in which a qubit is located on each edge of $L \times L$ square lattice
as shown in Fig.~\ref{fig2s}(a), 
is stabilized by the face and vertex stabilizer operators,
$A_f=\prod _{ i \in E_f } Z_{i}$ and 
$B_v=\prod _{j \in E_v} X_j$, respectively.
Here $Z_i$ and $X_j$ denote
Pauli operators on the $i$th and $j$th qubits,
and $E_{f}$ and $E_v$ indicate the sets of four edges 
surrounding a face $f$ and adjacent to a vertex $v$, respectively.
The error syndromes $\{ a_f\}$ and $\{ b_v\}$
are defined as sets of eigenvalues of the face and vertex stabilizers,
respectively,
which are used to identify $X$ and $Z$ errors.
These errors are assumed to occur on each qubit with independent and identical error probability $p$,
for simplicity.

In standard topological quantum error correction (STQEC) \cite{Dennis}, 
the error syndrome is extracted to the classical world
through parallel projective measurements.
According to the error syndrome, we use a classical algorithm, 
minimum-weight-perfect-matching (MWPM) \cite{Edmond},
which tells us the location of the errors.
By virtue of the locality and translational invariance
of the surface code,
the error threshold of the STQEC is very high $\sim 1\%$
even when using only the nearest-neighbor two-qubit gates
to extract the error syndrome~\cite{Raussendorf06,Raussendorf07a,Raussendorf07b}.
However, the necessity of 
parallel projective measurements and the classical processing
may limit the effectiveness of STQEC.

If measurement-free QEC is allowed,  
both of these limiting factors can be eliminated in fault-tolerant quantum computation.
In Refs. \cite{Fitzsimons09,PazSilva11},
measurement-free QEC with a non-topological code was investigated,
where errors are corrected by unitary dynamics
with selective addressability of the boundary qubits.
It is well known that the projective measurements 
can be replaced with preparations of fresh ancillae (by dissipation)
followed by the controlled-unitary operations.
However,
if we straightforwardly apply this strategy to the surface code,
the classical processing, i.e., MWPM,
is too complex to be implemented by unitary dynamics.
The unitary dynamics for MWPM is far from local and translationally invariant, which completely diminishes the merit of using the surface code.
It is nontrivial to design the system to be topologically protected
with local and translationally invariant physical operations.
The proposed scheme, MFTP, makes active use of dissipative dynamics  
to reduce the entropy of the quantum system in such a way that
quantum information is topologically protected
with local and translationally invariant operations,
thus enabling us to fully utilize the advantage of 
the surface code.

\section*{Discrete-dissipative feedback for topological protection}
Recently, extensive research has been conducted
on engineering of dissipative dynamics to simulate open quantum systems~\cite{Muller} or 
to prepare quantum states of interest~\cite{Kraus,Verstraete,Vollbrecht11,Pastawski}.
Unfortunately, 
the local and translationally invariant dissipations
toward thermal equilibrium at finite temperature cannot protect
quantum information
due to no-go theorems of 
self-correcting quantum memory~\cite{BravyiTerhal,Beni}.
While extensive effort is made such as 
the toric-boson model~\cite{Hamma},
self-correcting quantum memory has not been achieved
by dissipative dynamics under a fixed local Hamiltonian system.
A quantum version of a non-volatile magnetic storage device in equilibrium,
such as a hard disk drive, seems to be hard to achieve at finite temperature.
Not only dissipative dynamics but also 
time dependence of the Hamiltonian,
which drives the system into nonequilibrium
with a feedback mechanism, would be a key ingredient to achieve topological protection.
In this sense, the proposed model achieving topological protection
of quantum information in nonequilibrium can be 
regarded as a quantum analog of dynamic memory,
which periodically refreshes the capacitor charge to store information reliably.

\begin{figure*}
\begin{center}
\includegraphics[width=150mm]{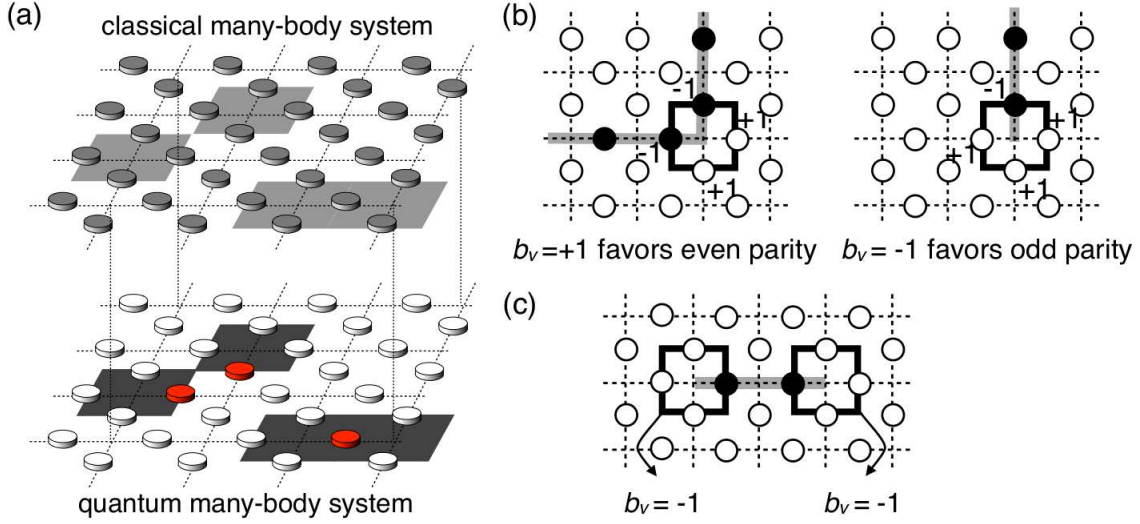}
\caption{The system and dissipative dynamics for MFTP. (a) The quantum (lower layer) and classical (upper layer) many-body systems for quantum information storage and its feedback controls, respectively. The $Z$ errors occur on the qubits colored by red. The associated vertex stabilizers with the eigenvalue $-1$ are denoted by black squares.
The four classical spins adjacent to each vertex interact according to the eigenvalues of the vertex stabilizers. The wrong-sign plaquettes with $b_v=-1$
are shown by gray squares on the upper layer. 
The double layer is not necessarily required, but
different degrees of freedom of the particles located on one layer
can be employed. 
(b) The effect of plaquette interaction.
A correct-sign plaquette with $b_v=+1$ favors an even number of $-1$s, where a chain of $-1$s is energetically
hard to terminate. The wrong-sign plaquette
favors an odd number of $-1$s, where a chain of $-1$s terminates easily.
(c) The competing interactions reveal the location of errors:
two wrong-sing plaquettes are connected with a short length.
}
\label{fig2s}
\end{center}
\end{figure*}

We utilize a  classical 2D many-body system, in addition to the quantum system,
as an ancilla to allow a feedback operation of a discrete type
(the upper layer in in Fig.~\ref{fig2s} (a)).
For simplicity, we explain the feedback operation 
for the $Z$ error correction only.
(The $X$ error correction can also be done in a similar way.)
According to the error syndrome $\{b_v\}$,
we cool (dissipate the energy) the classical system down to a sufficiently low 
but a finite temperature $T$
under a Hamiltonian 
\begin{eqnarray}
H (\{ b_v \})= -   J \sum _{v} b_v \prod _{i \in E_v}  u_i  - h \sum _i u _i,
\label{cooling_Hami}
\end{eqnarray}
where $u_i=\pm 1$ denotes the classical spin located 
on edge $i$, and $J$ and $h$ indicate the coupling constant and 
the magnetic field strength, respectively.
Here the ``classical" spins refers to the qubits
without their phase coherence. 
We call this Hamiltonian the 2D random-plaquette gauge model (RPGM) with magnetic fields.
Then the equilibrium configuration is obtained and
put back into the quantum system by transversal controlled-$Z$
(CZ) operations between the classical (controls) and quantum (targets) systems.

The above discrete-dissipative feedback process succeeds in correcting errors 
in the surface code through the following progression.
The location of errors in the surface code
can be identified by finding a minimum path connecting 
pairs of incorrect eigenvalues $b_v=-1$~\cite{Dennis}.
The first term in Eq.~(\ref{cooling_Hami}), 
which we call the plaquette interactions,
imposes the boundary conditions of a chain of $-1$s as shown in Fig.~\ref{fig2s} (b).
A correct-sign plaquette with $b_v=1$ energetically favors an even number of $-1$s,
where a chain of $-1$s is hard to terminate.
A wrong-sign plaquette with $b_v=-1$
energetically favors an odd number of $-1$s, where
a chain of $-1$s terminates easily.
On the other hand, the second term in Eq.~(\ref{cooling_Hami}), 
which we call magnetic fields, favors configurations of fewer $-1$s.
If the plaquette interactions 
are chosen to be strong enough to connect pairs of wrong-sign plaquettes
in the presence of the magnetic fields, 
these two competing interactions are expected to reveal 
the locations of the $Z$ errors at a sufficiently low temperature
(see Fig.~\ref{fig2s} (c)).
In Appendix A, we show that,
by choosing the strength of plaquette interactions  such that $4J/(2h) =\alpha \log L $ 
with a constant $\alpha$,
the logical error probability per step
decreases polynomially in the system size $L$,
as long as the physical error probability $p$ is below a threshold value.
Since the correlation length of the system scales like $O(\log L)$,
the relaxation time is expected to depend on the system size $L$
polylogarithmically (See Appendix A).
 
\begin{figure*}
\begin{center}
\includegraphics[width=150mm]{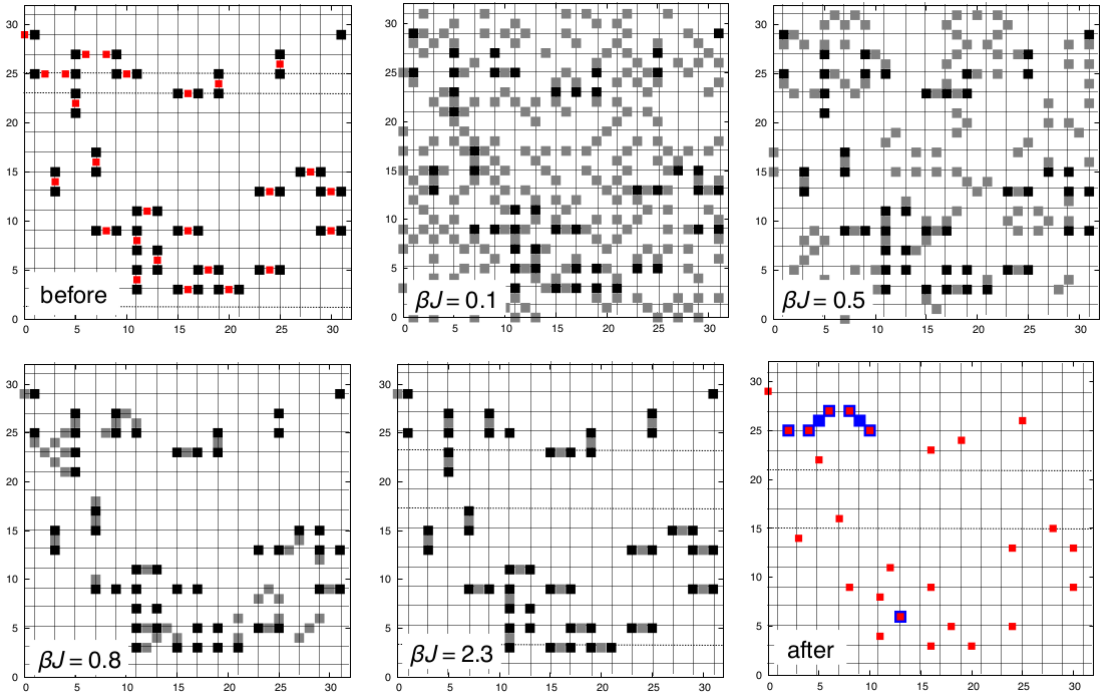}
\caption{Cooling dynamics and topological error correction.  
The location of the $Z$ errors in the quantum system
and wrong-sign plaquettes are shown by red and black squares, respectively (top left).
Equilibrium configurations are shown for $\beta J=0.1$ (top middle), 
$0.5$ (top right), $0.8$ (bottom left), and $2.3$ (bottom middle),
where $-1$ spins are shown by gray squares.
An equilibrium configuration at low temperature is 
put back into the quantum system. The remaining $Z$ errors 
after the feedback is shown by blue squares (bottom right).
Most of the errors are corrected, but some errors,
shown by blue squares, remain.
These are corrected by subsequent cycles.
}
\label{fig3s}
\end{center}
\end{figure*}
In order to confirm the above observation,
we performed numerical simulations of
the classical cooling process under the Hamiltonian $H( \{ b_v\})$
using the Metropolis method.
Figure ~\ref{fig3s} illustrates equilibrium configurations 
at $\beta h =0.1$, $0.5$, $0.8$, and $2.3$,
where $\beta$ is the inverse temperature, and $J=h$ is adopted, for example.
The remaining $Z$ errors after the feedback operation
are also shown.
Most of the errors are corrected within one cycle of MFTP
as shown in Fig.~\ref{fig3s} (lower right panel).
However, some errors still remain,
which are attributed to the excitations in the plaquette interactions
that result from a finite temperature effect.
Such errors are corrected in the following MFTP cycles
and/or suppressed by the logarithmic scaling of the plaquette interactions $4J/(2h)= \alpha  \log L$.

\begin{figure*}
\begin{center}
\includegraphics[width=150mm]{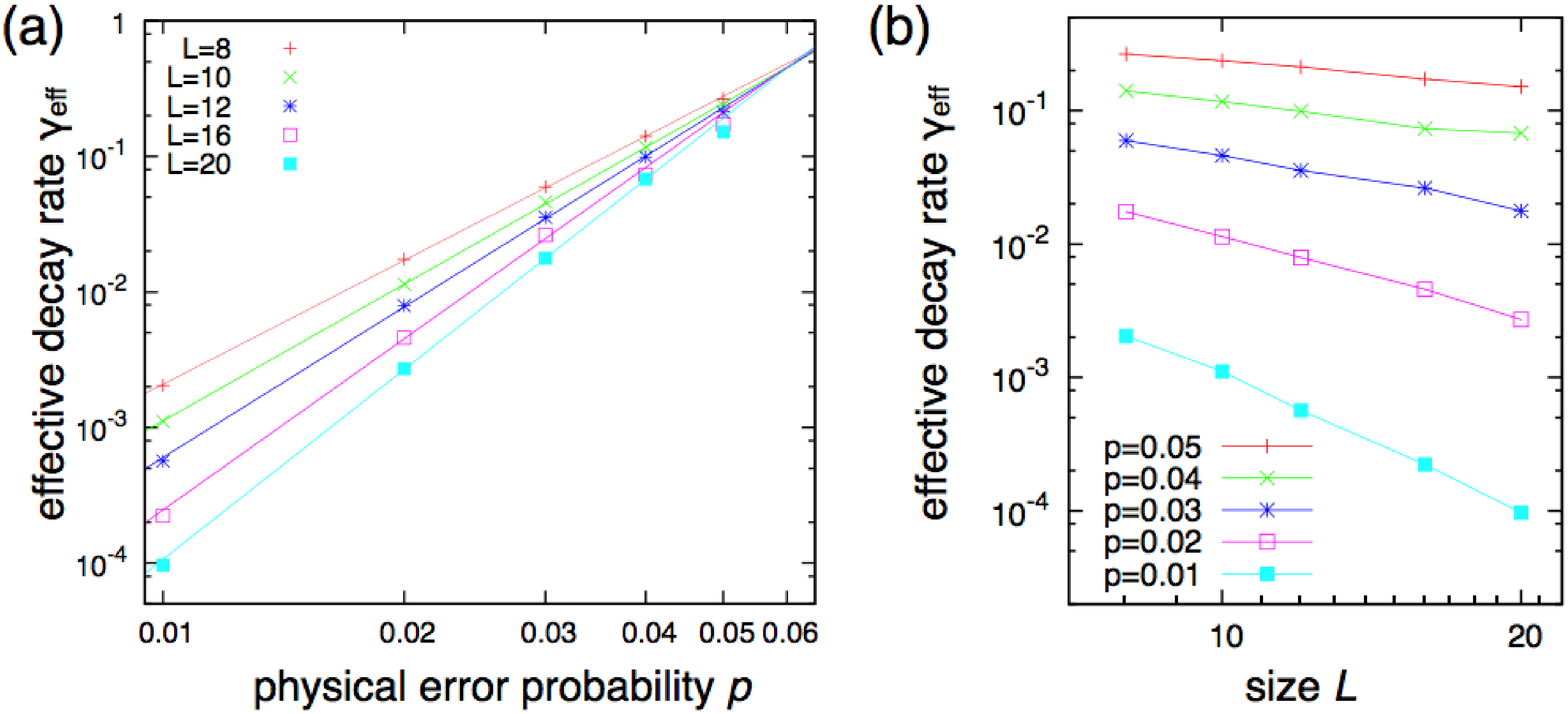}
\caption{Numerical results. 
The effective decay rate $\gamma _{\rm eff}$
is estimated by numerical simulations
for the physical error probabilities $p=0.01, 0.02,...,0.05$
and the system size $L=8,10,12,16,20$.
(a) The effective decay rate $\gamma _{\rm eff}$ as a function of the physical error probability $p$.
From top to bottom $L=8,10,12,16,20$.
(b) The effective decay rate $\gamma _{\rm eff}$ as a function of the system size $L$.
From bottom to top $p=0.01,0.02,0.03,0.04,0.05$.
}
\label{Numerical}
\end{center}
\end{figure*}
We have further investigated 
the logical error probability 
with varying the system size $L$ and the physical error probability $p$.
Specifically we have chosen $\alpha =1$ and utilized 
thermal equilibrium state of a temperature 
$T=1/\beta  = -2h/ \ln [p/(1-p)]$ on the Nishimori line~\cite{Nishimori},
where the thermal fluctuation and physical error probability are balanced.
In the limit of large $J$ (i.e. large $\alpha$), the discrete-dissipative feedback
under the Nishimori temperature
achieves an optimal decoding of the surface code.
The effective decay rate $\gamma _{\rm eff}$ of the stored
information under topological protection
is calculated by fitting the logical error probability 
to $3(1-e^{ - \gamma _{\rm eff} t})/4$ with $t$ being the time step.
The effective decay rate $\gamma _{\rm eff}$ is plotted as functions of 
the physical error probability $p$ and the size $L$
in Fig. \ref{Numerical} (a) and (b), respectively.
We have observed that 
the threshold for physical error probability with $\alpha = 1$ is 
at least as high as $5.0\%$, and the extrapolation implies 
that the threshold lies around $6.0\%$.
We have also confirmed that the logical error probability decreases rapidly by increasing the system size $L$.
 
\section*{Implementation and feasibility of MFTP} 
Next we consider a physical implementation of MFTP.
\begin{figure*}
\centering
\includegraphics[width=120mm]{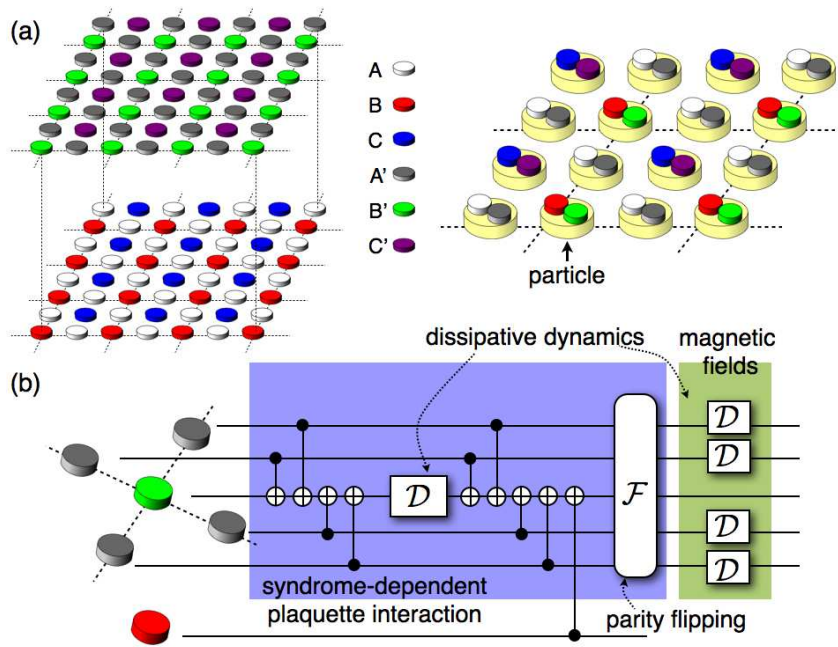}
\caption{Physical implementation of MFTP. 
(a) The double-layer system for MFTP. Each 2D layer consists of three 
species of particles (left). 
The different degrees of freedom of the particles on a single 2D layer can also be utilized (right). (b) The stabilizer pumping scheme
for a digital simulation of the cooling process (see Appendix B).}
\label{fig5s}
\end{figure*}
We consider a double-layer system, each of which consists of three species of particles as 
shown in Fig.~\ref{fig5s} (a) (left). 
The A qubits and A' spins are the quantum and classical systems
considered, respectively.
The  B and C qubits are 
the ancillae for the $Z$ and $X$ error syndrome extractions, respectively.
The B' and C' spins are used to mediate the 
syndrome-dependent plaquette interactions.
Instead of the double-layer system,
particles on a single layer, which have different degrees of freedom,
can be utilized as shown in Fig.~\ref{fig5s} (a) (right).

We utilize simultaneous two-qubit gates
on neighboring particles of different species,
such as controlled-NOT (CNOT) gates between A and B qubits, for example. In addition, we use a dissipative operation,
specifically T$_1$ relaxation or incoherent pumping, on the classical spins.
The procedure in each MFTP cycle for the $Z$ error correction is as follows (see also Fig.~\ref{MFTPcycle}):
\begin{figure*}
\centering
\includegraphics[width=150mm]{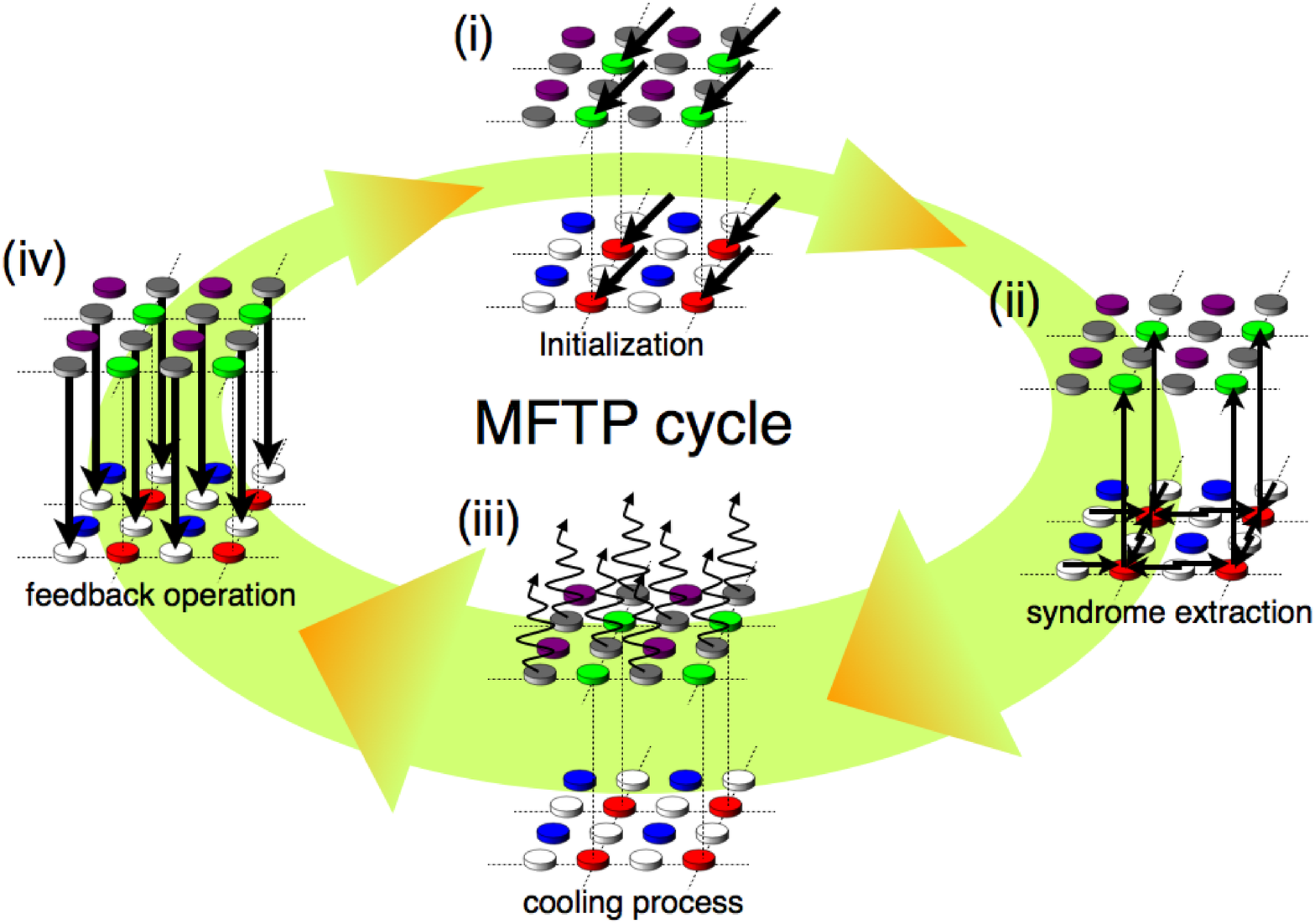}
\caption{One MFTP cycle for the $Z$ error correction. (i) The B qubits and B' spins
are initialized to extract and copy the error syndrome, respectively. (ii) The error syndrome 
is extracted from the A qubits to the B qubits. The extracted error syndrome is copied to the B' spins.
(iii) By using the copied error syndrome, two-qubit gates, and single-qubit dissipative dynamics,
the cooling process under the Hamiltonian $H(\{b_v\} )$ is simulated in a digitalized way (see Fig.~\ref{fig5s} (b)).
(iv) The obtained equilibrium configuration, parts of which reveal the location of the $Z$ errors,
is put back into the A qubits to correct the $Z$ errors.
}
\label{MFTPcycle}
\end{figure*}

\begin{enumerate}[(i)]
\item Initialize B and B' to $|0\rangle$. 

\item Perform CNOT gate operations between A (controls) and B (targets)
to extract the error syndrome.
Perform CNOT gate operations between B (controls) and B' (targets) to 
copy the syndrome.
\label{syn_ex}

\item Simulate the cooling dynamics in a digitalized way
by using the stabilizer pumping \cite{Weimer,Barreiro,Lanyon,Muller} (see Fig.~\ref{fig5s} (b)).
\label{H_P}

\item Perform C$Z$ gate operations between A' (controls) and A (targets)
to apply the feedback operation.
\end{enumerate}
A similar procedure for the $X$ error correction is 
also performed with a basis change using Hadamard operations.
These MFTP cycles for the $X$ and $Z$ error corrections
are repeated.

The cooling process in step (iii)
is simulated in a digitalized way,
which we call digitalized cooling,
using the stabilizer pumping \cite{Weimer,Barreiro,Lanyon,Muller}.
We can realize the syndrome-dependent plaquette interactions automatically
by using the five-body interactions described by
\begin{eqnarray}
\tilde H = -J \sum _{v} Z_v^{B'} \prod _{i \in E_v} Z_i^{A'}  - h \sum _{i}Z_i^{A'},
\end{eqnarray}
since the error syndrome is copied onto the B' spins in step (ii).
Note that the classical spins A', B', and C' 
are denoted as if they are qubits for simplicity of the notation.
By introducing an interaction $H_{\rm int}$ 
between the ancilla classical spins and the environment,
the time evolution of the total system is given by $V(t)\equiv e^{-i (\tilde H + H_{\rm int})t}$.
If the coupling strength with the environment is sufficiently small, 
and the correlation time is sufficiently short, then
the time-evolution of the system can be regarded as a Markovian decay.
Then, the time evolution is divided into short digitalized cooling steps 
of interval $\tau = t/m$:
\begin{eqnarray*}
V(t) \simeq [e^{- i (H_P + H_{\rm int}/2) \tau } e^{- i (H_F + H_{\rm int}/2) \tau} ]^{m},
\end{eqnarray*}
where $H_P\equiv -J \sum _{v} Z^{B'}_v \prod _{i \in E_v} Z^{A'}_i$, 
$H_F\equiv- h \sum _{i}Z_i^{A'}$, 
and the Trotter-Suzuki expansion~\cite{Suzuki} is employed.
By tracing out the environment,
we obtain Markovian digitalized cooling dynamics 
\begin{eqnarray*}
 {\rm Tr} _{E} [V(t) \rho _{A} \otimes  \rho _{E} V(t) ^{\dag}]
\simeq [ e^{\tau \mathcal{L}_P} e^ {\tau \mathcal{L}_F} ] ^{ m} \rho _A
\end{eqnarray*}
where $\rho _A$ and $\rho _E$ indicate density matrices of the ancilla system and the environment, respectively.
$\mathcal{L}_P$ and $\mathcal{L}_F$ denote
the Lindblad superoperators~\cite{Lindblad} with respect to the Markovian decays under 
$H_P$ and $H_F$, respectively.
These are
realized by using single-qubit dissipative dynamics and two-qubit gates
as shown in Fig.~\ref{fig5s} \cite{Weimer,Barreiro,Lanyon,Muller} (see Appendix B).

Let us discuss requirements on the physical parameters in experiments.
The decoherence rate of the quantum system A, B, and C is denoted by $\Gamma$.
The strength of the dissipative operation on the classical spins  A', B',  and C'
is denoted by $\gamma \sim || H_{\rm int} ||$.
We also define a coupling 
strength $\kappa$ between the qubits and spins, 
which limits the gate time $1/\kappa$. 
Furthermore, we utilize a Markov approximation, $J,h \gg \gamma $,
and the Trotter-Suzuki expansion imposes
$J\gamma \tau^2 \ll 1$ and $h \gamma \tau ^2 \ll 1$.
By setting $J,h \sim 10\gamma$, $J\gamma \tau ^2 \sim 10^{-1}$
and $h\gamma \tau^2 \sim 10^{-1}$, 
the time interval $\tau$ of each digitalized cooling step becomes
$\tau \sim 10^{-1}/\gamma$.
According to the numerical simulations, 
the cooling process takes 
$\sim 10^{2}$-$10^{4}$ Monte Carlo steps,
each of which physically takes a relaxation time of $\sim 1/\gamma$
for local spin flipping.
Thus the cooling time $t_{\rm cool}$ required for one MFTP cycle is estimated to be $t_{\rm cool} \sim 10^2$-$10^4/\gamma$.

Since the cooling time $t_{\rm cool}$ is divided into digitalized cooling steps 
of time interval $\tau \sim 10^{-1}/\gamma$,
the number of repetitions is $m = t_{\rm cool} /\tau \sim  10^3\textrm{-}10^5$.
In each digitalized cooling step, only a few unitary operations are required
since most of the unitary gates are commutable and operate simultaneously.
Accordingly, in addition to the cooling time $t_{\rm cool}$,
the unitary gate operations take $\sim m/g$.
As a result, the time taken by
one MFTP cycle is
\begin{eqnarray*}
t_{\rm cycle} =t_{\rm cool} +m/\kappa  \sim 10^{2} /\gamma + 10^{3}/\kappa .
\end{eqnarray*}
In order for MFTP to work,
the physical error probability $ p= 1- e^{ -  \Gamma  t_{\rm cycle}} $ for the quantum system have to be as small as $10^{-2}$.
Hence, we obtain requirements on the physical parameters,
$\Gamma /\gamma  \lesssim  10^{-4}\textrm{-}10^{-6}$ and  $\Gamma /\kappa \lesssim 10^{-5}$.
The former and latter conditions indicate that 
the speeds of dissipation and unitary operations have 
to be $10^4\textrm{-}10^6$ and $10^{5}$ times faster than
the decoherence time of the quantum system,
respectively. 

Let us discuss the feasibility of the above conditions
in the case of a solid state system with nuclear and electron spins
for quantum and classical systems, respectively.
State-of-the-art chemistry and nanotechnology 
research has demonstrated the possibility of
engineering materials 
with nuclear and electron spins arranged on a lattice.
In solids like diamond, the T$_2$ relaxation time of nuclear spins exceeds 1 sec 
even if it is located near an electron spin~\cite{Negoro1}.
The electron spins can be initialized 
within $1/\gamma\sim 100$ nsec 
if they have appropriate optical transitions.
The hyperfine interaction between 
electron and nuclear spins and 
the dipolar interaction between electron spins
for two-qubit gates are both typically in the order of $\kappa\sim$1-100 MHz.
These parameters fulfill the above conditions.
Other degrees of freedom in natural or artificial atoms,
for example, as the motional and internal degrees of freedom in trapped atoms \cite{Muller,Barreiro,Weimer,Lanyon}
or superconducting flux qubits coupled with microwave cavities \cite{Fowler,Ghosh},
can also be utilized.

\section*{Conclusion}
We have proposed MFTP, 
which is a novel way to protect quantum information
from decoherence using dissipative operations
without parallel projective measurements
or selective addressing.
It has been shown that local and translationally invariant physical operations
without any selective addressing and measurements
achieve topological protection of quantum information.
The requirements of MFTP are feasible 
in various physical implementations.
Furthermore, the dissipative dynamics used in MFTP
is well studied in quantum simulation. 
Actually, the 
key ingredients of MFTP have already been demonstrated in quantum simulation~\cite{Weimer,Barreiro,Lanyon},
thus the proposed model establishes a promising means of progressing from 
quantum simulation to fault-tolerant quantum computing.

\section*{Acknowledgments}
The authors thank S.~Nishida, Y.~Morita, H. Tasaki and B.~Yoshida for valuable dicussions. 
This work was supported by the Funding Program for World-Leading Innovative Research \& Development on Science and Technology (FIRST), MEXT Grant-in-Aid for Scientific Research on Innovative Areas 20104003 and 21102008, the MEXT Global COE Program, JSPS Grant-in-Aid for Research Activity Start-up 25887034.

\appendix
\section*{Appendix}
\subsection*{A. Analysis of the logical error probability}
The energy penalty for the boundary mismatching
is $4J$, since they are always created 
as a pair of excitations with respect to plaquette interactions.
On the other hand,
a connected chain of $-1$s of a length $r$ 
costs an energy $2hr$ due to the magnetic fields.
Thus the pair of two wrong-sign plaquettes are connected in a typical case,
when the distance between two wrong-sign plaquettes 
is shorter than $r_{\rm cor} \equiv 4J/(2h)$,
which gives a typical length of the system
(more precisely the maximum correlation length at zero temperature).
If errors occur contiguously, and 
the distance between two wrong-sign plaquettes
becomes larger than $r_{\rm cor}$, then
the boundary mismatching is energetically favored
due to the energy penalty of magnetic fields.
This means that the spins of the classical system search
pairs of wrong-sign plaquettes inside an area 
whose vertical and horizontal lengths typically are given by $r_{\rm cor}$.

In the following, we will bound the logical error probability 
after the feedback operation through a two-step 
argument.
At first, we will consider a necessary condition: 
the energy landscape of the 
classical system $H(\{ b _v\})$ is given appropriately
such that the spin configuration at zero temperature
can correct the error appropriately.
This is not sufficient for our purpose, since
we utilize a spin configuration at finite temperature.
Thus, secondary, 
we will bound the logical error probability
for the given appropriate energy landscape 
taking into account the finite temperature effect.

For a given error chain $C_e$,
let us define a minimum-path chain $C_m$,
which connects pairs of wrong-sign plaquettes with a shortest path
(i.e. the solution of MWPM).
When errors are dense, the error and minimum-path chains $C_e + C_m$ would constitute a connected chain of a length longer than $r_{\rm cor}$.
In such a case, 
the error correction fails even if we have the spin configuration 
at zero temperature,
resulting in an inappropriate energy landscape.
To calculate such a failure probability $p_{\rm error}$,
let us consider an area whose
vertical and horizontal lengths are given by $r_{\rm cor}$.
The probability that 
the error and minimum-path chains $C_e + C_m$ 
constitute a connected chain 
of length $l \geq r_{\rm cor}$ can be calculated 
by following the method provided in
Ref. \cite{Dennis}:
\begin{eqnarray*}
p_{\rm error} <  \sum _{l \geq r_{\rm cor}} C(l) 
\sum _{k>l/2}^{l} \left( 
\begin{array}{c}
l
\\
k
\end{array}
\right)
p^{k} (1-p)^{l-k}
<
 \sum _{l \geq r} C(l) 2^{l} p^{l/2},
\end{eqnarray*}
where $C(l)$ is the number of self-avoiding walks
of length $l$ inside this area,
which is bounded as $C(l) < 3\times 4^{l-1}$.
Since we can begin such a connected chain of length $l \geq r_{\rm cor}$ 
at any one of $L^2$ lattice sites,
the total failure probability is at most $1-(1-p_{\rm error} )^{ L^2} \simeq
L^2 p_{\rm error} $.
(In STQEC with MWPM,
$r_{\rm error}$ is replaced with the system size $L$.
This reproduces the calculation made in Ref.~\cite{Dennis},
leading to the threshold value independent of the system size.
The fact that $r_{\rm error} = O(L)$ implies that communication
over the whole system is employed in the classical 
processing for MWPM,
which requires a time $O(L)$ in the presence of parallelism. 
By using an elaborated renormalization 
algorithm, we can reduce the time complexity for this communication
to $O(\log L)$~\cite{Poulin},
where non-local operations,
which are prohibited in the present setup, are still required.
See also Tab. \ref{table1}.)

Since we utilize a spin configuration at finite temperature,
we further take into account
excitations originated from the finite temperature effect
to calculate the logical error rate.
There are two types of excitations:
(i) excitations without boundary mismatches,
and (ii) excitations with boundary mismatches.
In the former case (i),
while the ground state configuration corresponds to 
an appropriate location of the errors,
excitations with respect to the magnetic fields 
help to form
a connected chain of length longer than $r_{\rm cor}$.
Such a probability is calculated as
\begin{eqnarray*}
p_{\rm ex} &=&  \sum _{l \geq r_{\rm cor}} C(l) 
\sum _{k=0}^{l/2} \left( 
\begin{array}{c}
l
\\
k
\end{array}
\right)
p^{k} (1-p)^{l-k} e^{-2\beta h(l-2k)}p_{\rm gs}
\\
&<&
  \sum _{l \geq r_{\rm cor}} C(l) 
\sum _{k=0}^{l/2} \left( 
\begin{array}{c}
l
\\
k
\end{array}
\right)
[e^{-2\beta h} (1-p)]^{l-k} (p/e^{-2\beta h })^k
\\
&<&
  \sum _{l \geq r_{\rm cor}} C(l) 
 2^{l}p^{l/2} 
\end{eqnarray*}
where $p_{\rm gs} \geq 1$ is the population of 
the ground state,
and the temperature is chosen to be $e^{ - 2 \beta h}=p/(1-p)$.
The condition $e^{ - 2 \beta h}=p/(1-p)$,
so-called Nishimori line \cite{Nishimori}, indicates 
that the thermal fluctuation with respect to the magnetic field
and the physical error probability are balanced.
As a result, we can obtain the same scaling of 
the failure probabilities $p_{\rm error}$
and $p_{\rm ex}$
to the leading order in the presence of the thermal fluctuation below the Nishimori line.
Similarly to the previous case,
the connected chain of length $l \geq r_{\rm cor}$ 
can begin at any one of $L^2$ lattice sites,
and hence the total failure probability in this case 
is at most $L^2 p_{\rm ex}$.

Finally, we consider the latter case (ii).
Since the number of connected errors in this area is 
at most $r_{\rm cor}/2$,
the boundary mismatching costs
at least $4J - 2h(r_{\rm cor}/2)$ energy penalty.
Thus the probability of the excitation 
causing the boundary mismatching 
can be given by
\begin{eqnarray*}
p_{\rm b.m.} = e^{ -4\beta J + 2\beta h(r_{\rm cor}/2) }p_{gs}
< e^{-\beta hr_{\rm cor}} .
\end{eqnarray*}
Such a boundary mismatching can occur 
at any one of $L^2$ plaquettes,
the total failure probability in this case
amounts $L^2 p_{\rm b.m.}$.

Summing all these failure probabilities,
the logical error probability 
is given by
\begin{eqnarray*}
p_{\rm logi} = L^2 p_{\rm error} + L^2 p_{\rm ex} + L^2 p_{\rm b.m.}
< L^2   \left( \sum _{l \geq r_{\rm cor}} 
\frac{8}{3} 6^{l}p^{l/2}  + [p/(1-p)]^{r_{\rm cor}/2} \right).
\end{eqnarray*}
Suppose the typical length $r_{\rm cor}$ 
is chosen to be $r_{\rm cor} = \alpha \log L$
with a constant $\alpha$.
The failure probability converges to zero
in the limit of a large $L$
if $p/(1-p)\leq e^{- 4/\alpha}/36$. 
For example, with $\alpha =1$ and $2$,
this condition reads the threshold value $p_{\rm th} \simeq 5.1 \times 10^{-4}$ and $3.7 \times 10^{-3}$,
respectively.
We should note that the above calculation
is far from tight,
since $C(l)$ is overestimated substantially.
Furthermore in many cases 
the boundary mismatches do not cause a fatal failure,
since the remaining errors would be corrected in the following steps.
Thus the true threshold value is expected to be 
much more higher as shown in the numerical simulation.

The relaxation time is an important factor for the proposed scheme,
since the physical error probability $p$ is increased exponentially
during the delay of the feedback.
The relaxation time is typically given as a polynomial function of the correlation length 
of the system.
(In a critical system, 
the typical time scale of the system $\tau _{\rm cri}$ and the correlation 
length $r_{\rm cri}$ are related by $\tau _{\rm cri} = r_{\rm cri}^{z}$
with $z$ being the dynamical critical exponent.
The present classical system $H(\{b_v\})$ is not 
critical, the relaxation time is expected to be 
a lower order polynomial function than $r_{\rm cor}^{z}$.)
In the present case, 
the typical length of the system 
depends on the system size only logarithmically, $r_{\rm cor} = \alpha \log  L$.
Thus the relaxation time required is expected 
to be a polylogarithmic function of the system size $L$.
Then, the delay time $\tau$ for the feedback 
scales like $ \tau \sim {\rm polylog}(L)$.
The threshold for the decoherence rate of the quantum system 
scales like $\Gamma _{\rm th} \sim 1/{\rm polylog}(L)$.

Let us finally consider a large $J$ limit,
where $J$ is scaled as $O(hL)$.
In such a case, $r_{\rm cor}$ scales like $O(L)$.
Thus the discrete-dissipative feedback at the Nishimori temperature 
provides an optimal decoding,
which is slightly better than the decoding with MWPM~\cite{Ohzeki,Merz}.
An exponential suppression of the logical error 
can also be achieved under this scaling.
However, the coupling $J \sim O(hL)$ seems to be too strong 
for the proposed scheme to be interesting,
since it takes a long time to obtain the spin configuration.
We summarize the performance of the proposed scheme 
comparing with other related models 
including decoding methods
based on the explicit parallel measurements
and the classical processing in Table \ref{table1}.
\begin{table}[t]
{\footnotesize
\begin{center}
\begin{tabular}{c||c|c|c|c|c}
\hline
model  & \shortstack{coupling \\
strength} & \shortstack{memory  \\ time} & \shortstack{decoding \\ time}  & \shortstack{type of operations \\ employed (in 3D)} & dynamics
\\
 \hline \hline
\shortstack{our proposal \\
 with $J = O(h\log L)$} &
$O(\log L)$ &
 ${\rm poly}(L)$ & ${\rm polylog}(L)$ & \shortstack{local and \\ translationally invariant} & \shortstack{nonequilibrium \\ driven by feedback}
\\
\hline
\shortstack{our proposal \\
with $J = O( hL)$} &$O(L)$ 
 & ${\rm exp}(L)$ & ${\rm poly}(L)$ & \shortstack{ local and \\ translationally invariant} & \shortstack{nonequilibrium \\ driven by feedback}
\\
\hline
\shortstack{surface code in 4D \\ 
w/o error correction} & {\rm const.} &
${\rm exp}(L)$ & no & \shortstack{non-local \\ translationally invariant} & thermal equilibrium
\\
\hline
\shortstack{
surface code in 2D
\\
w/o error correction} & 
const.  &
const. &  no  & \shortstack{ local and \\ translationally invariant} & thermal equilibrium
\\
\hline
toric-boson model~\cite{Hamma} & 
${\rm poly}(L)$ &
${\rm poly}(L)$ & no & \shortstack{ local and \\ translationally invariant} & thermal equilibrium
\\
\hline
\shortstack{surface code in 2D
\\
with MWPM} &
\shortstack{not \\ defined} & 
 ${\rm exp}(L)$ & ${\rm poly}(L)$ & \shortstack{non-local and \\ no translationally invariant}  & \shortstack{nonequilibrium \\ driven by feedback}
\\
\hline
\shortstack{
surface code in 2D
\\
with a renormalization
\\ decoder~\cite{Poulin}} 
&\shortstack{not \\ defined} & ${\rm exp}(L)$ & ${\rm polylog}(L)$ & 
\shortstack{non-local but \\ having self-similarity} & \shortstack{nonequilibrium \\ driven by feedback}
\\
\hline
\end{tabular}
\end{center}
}
\caption{A comparison of the performances of the related  topological protection schemes based on the surface codes. 
}
\label{table1}
\end{table}%

\subsection*{B. Digital simulation of the cooling process}
The Lindblad super-operators $\mathcal{L}_P$ and
$\mathcal{L}_{F}$ for Markovian decays
under the Hamiltonian $H_P$ and $H_F$ are given by
\begin{eqnarray}
\mathcal{L}_P \rho &=& -\gamma _{-} \sum _{v \in B'} \left( \hat{\eta}^{\dag}_v  \hat{\eta}_v \rho + \rho  \hat{\eta}_v ^{\dag} \hat{\eta}_v - 2  \hat{\eta}_v \rho  \hat{\eta}_v^{\dag} \right)
\nonumber
\\
&&-\gamma _{+} \sum _{v \in B'}  \left( \hat{\eta}_v \hat{\eta}^{\dag}_v \rho + \rho  \hat{\eta}_v  \hat{\eta}^{\dag}_v - 2  \hat{\eta}_v^{\dag} \rho  \hat{\eta}_v \right),
\label{H_P}
\\
\mathcal{L}_F \rho &=&
- \gamma' _{-} \sum _{i \in A'} \left(  \hat \sigma ^{+}_i \hat \sigma ^{-}_{i}  \rho + \rho \hat \sigma ^+_i \hat \sigma^-_i - 2 \hat \sigma^-_i \rho \hat \sigma^+_i \right)
\nonumber
\\
&& -\gamma' _{+} \sum _{i \in A'}  \left( \hat \sigma^-_i \hat \sigma^+_i \rho +   \rho \hat \sigma^-_i \hat \sigma^+_i - 2 \hat \sigma^+_i \rho \hat \sigma^- _i \right).
\label{H_F}
\end{eqnarray}
The operator $\hat{\eta}_v =   X^{A'}_{i \in E_v} (I - Z^{B'}_v \prod _{i \in E_v} Z^{A'}_i)/2$,
where $i \in E_v$ means that one of the edges in $E_v$ is chosen randomly, lowers
the energy with respect to $- J Z^{B'}_v \prod _{i \in E_v} Z^{A'} _i$.
The operators $\hat \sigma ^{+}_i = |1\rangle \langle 0|_i$ and $\hat \sigma^{-} _i =|0\rangle \langle 1|_i$
are the raising and lowering operators for the $i$th qubit, respectively.
The ratio of the decay and the excitation rates is given 
in terms of the inverse temperature of the environment
and the energy gaps $J$ and $h$ of the system as follows:
$\gamma _- / \gamma _+ = e^{ - 2\beta J}/(1+ e^{ - 2\beta J})$ and  $\gamma' _- / \gamma' _+ = e^{ - 2\beta h}/(1+ e^{ -2 \beta h})$.
The single-qubit dissipative dynamics
$\mathcal{L}_{F}$
is implemented 
using a T$_1$ relaxation or incoherent pumping.

For the realization of $\mathcal{L}_{P}$ through
single-qubit dissipative dynamics and two-qubit gate operations,
we introduce another lowering operator
$\hat \xi = X^{B'}_v (I - Z^{B'}_v \prod _{i \in E_v} Z^{A'}_i)/2$.
If the parity of $Z^{B'}_v  \prod _{i \in E_v} Z^{A'}_i$ is odd,
this operator flips the spin located at the vertex $v \in B'$, which is the copy of the syndrome $b_v$.
Our goal is reducing the energy of the classical system consisting of the A' spins
by appropriately flipping the parity of $\prod _{i \in E_v} Z^{A'}$ 
as done by $\hat {\eta}$.
To this end, if the copy of the syndrome is flipped,
we have to flip either one of the ancilla spins $i \in E_v$.
To establish whether the copy of the syndrome is flipped or not,
we apply a CNOT gate between the B qubits (controls) to the B' spins (targets).
Since the B' spins are copies of the B qubits,
this CNOT gate reveals whether the B' spins B' are flipped.
If a B' spin is not flipped,
the state is $+1$ after the application of the CNOT gate.
If a B' spin is flipped, the state is $-1$ after the CNOT gate.
Consequently, 
we can change the parity of 
the A' spins on the corresponding plaquette 
with random CNOT gates $\mathcal{F}$ between
the B' spins (controls) and A' spins (targets),
\begin{eqnarray*}
\mathcal{F} \rho = \frac{1}{4} \sum _{i \in E_v} \Lambda (X)_{v,i} \rho \Lambda (X)_{v,i},
\end{eqnarray*}
where $\Lambda (X)_{v,i}$ is the CNOT gate acting on
the $v$th and $i$th particles as the control and target, respectively.
To change the parity with unitary operations,
the random CNOT gates can be replaced with
the controlled-$\exp[{- i (\pi/4) X^{A'}}]$ operation, for example.
In this case, the parity is not changed with unit probability,
and hence it may slow down the cooling dynamics.
The CNOT gate from the B qubits  (controls) to the B' spins (targes)
and the parity flipping operation $\mathcal{F}$
are denoted together by $\tilde {\mathcal{F}}$.
Followed by the operation $\tilde {\mathcal{F}}$,
the effective action of the $\hat \xi$ is equivalent to $\hat \eta$.

Finally we describe the dissipative operation under $\hat \xi$,
whose Lindblad super-operator is denoted by
 $\tilde{\mathcal{L}}_{P}$.
 The lowering operator $\hat \eta$ is transformed 
 by $U\equiv \prod _{i \in E_v} \Lambda (X)_{i,v}$ (see Fig.~\ref{fig5s} (b)) to 
 \begin{eqnarray*}
 \hat \xi _v =  U X_v \left( I - Z_v \prod _{i \in E_v} Z_i \right) U = X_v ( I - Z_v )/2 = \sigma ^{-}_v, 
 \end{eqnarray*}
 which is the single-qubit lowering operator for the spins
and hence implemented using T$_1$ relaxation or incoherent pumping.
Thus, by combining the operation $\tilde {\mathcal{F}}$
and $\mathcal{U}$,
the dissipative operation ${\mathcal{L}_{P}}$ can 
be realized as
\begin{eqnarray*}
e^{ \tau { \mathcal{L}}_{P} } \rho = \tilde {  \mathcal{F}} \mathcal{U} e^{\tau \tilde {\mathcal{L} }_{P}} \rho,
\end{eqnarray*}
where $\mathcal{U}\rho \equiv U \rho U$.

\end{document}